\documentclass[multphys,vecphys]{svmult}

\usepackage{makeidx}   
\usepackage{graphicx}  
\usepackage{multicol}  
\usepackage{cite}
\usepackage{epsfig}
\makeindex             

\newcommand{\be}{\begin{equation}}
\newcommand{\ee}{\end{equation}}
\newcommand{\bey}{\begin{eqnarray}}
\newcommand{\eey}{\end{eqnarray}}
\newcommand{\<}{\langle}
\renewcommand{\>}{\rangle}
\def\Eq#1{Eq.~(\ref{#1})}
\def\Eqs#1#2{Eqs.~(\ref{#1}-\ref{#2})}

\def\tmod{|\tau|}
\def\thermalfield{\tau}

\def\thermalfieldzero{\thermalfield(0)}
\def\psizero{\psi(0)}
\def\hzero{h(0)}
\def\thermalfieldl{\thermalfield(l)}
\def\psil{\psi(l)}

\def\thermalfieldzero{\thermalfield}
\def\psizero{\psi_0}
\def\hzero{h}
\def\thermalfieldl{\thermalfield}
\def\psil{\psi}

\def\yT{y_\thermalfield}
\def\yTau{{y_\tau}}
\def\yH{{y_h}}
\def\ypsideux{y_{\psi^2}}
\def\ypsitrois{y_{\psi^3}}
\def\yTpsi{y_{\thermalfield\psi}}
\def\yHpsi{y_{h \psi}}

\begin{document}
\title*{Logarithmic corrections and universal amplitude ratios in the
    4-state Potts model}
\toctitle{Logarithmic corrections and universal amplitude ratios in the
    4-state Potts model}
%
\titlerunning{Logarithmic corrections in the
    4-state Potts model}
%
\author{
B. Berche\inst{1},
P. Butera\inst{2},
L.N. Shchur\inst{3}}
\authorrunning{B. Berche et al.}
%
%
\institute{
Laboratoire de Physique des Mat\'eriaux,
Universit\'e Henri Poincar\'e, Nancy I, BP 239,
F-54506 Vand\oe uvre les Nancy Cedex, France\\
\texttt{berche@lpm.u-nancy.fr}
\and
Istituto Nazionale di Fisica Nucleare,
Universit\'a Milano-Bicocca,
Piazza delle Scienze 3, 20126, Milano, Italia\\
\texttt{paolo.butera@mib.infn.it}
\and
Landau Institute for Theoretical Physics,
Russian Academy of Sciences,
Chernogolovka 142432, Russia\\
\texttt{lev@landau.ac.ru}
}
\maketitle              

\begin{abstract}
    Monte Carlo and series expansion  data
    for the energy, specific heat, magnetisation and susceptibility of
    the 4-state Potts model in the vicinity of the
    critical point are analysed.
    The role of logarithmic corrections is discussed.
    Estimates of universal ratios $A_+/A_-$,
    $\Gamma_+/\Gamma_L$, $\Gamma_T/\Gamma_L$ and $R_c^+$ are given.
\end{abstract}

\today

\section{Introduction}
\label{sec:1}  
\begin{figure}[t]
\begin{center}
\includegraphics[angle=0,scale=0.40]{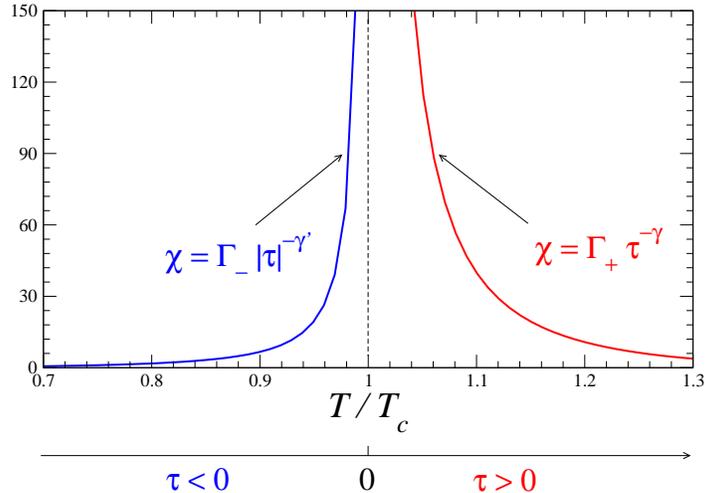}
\end{center}\vspace*{-0.4cm}
\caption{\label{fig:ChiMain}
Typical behaviour of the susceptibility at a second order phase transition.
The quantities $\gamma=\gamma'$ and $\Gamma_+/\Gamma_-$ are universal.}
\end{figure}

The study of critical phenomena and phase transitions is a
traditional subject of statistical physics which has known its
``modern age'', since powerful approaches have been developed
(renormalization group, conformal invariance, sophisticated
simulation algorithms, \dots). Simplified models attracted a lot of
attention. This is essentially due to a spectacular property of
continuous phase transitions at their critical point, {\em scale
invariance}, which leads to an extreme robustness of some
quantities, like the critical exponents which are thus referred to
as {\em universal quantities}. Only very general properties (space
or spin dimension, symmetry, range of interaction, \dots) determine
the universality class. This makes the theory of critical phenomena
a very efficient and predictive tool: As soon as one knows the
general characteristics of a physical system from general symmetry
arguments, it is possible in principle to predict exactly the
``shape'' of the singularities which are developed at the critical
point. The term ``exact'' is here understood rigorously, for example
a two-dimensional system with  the symmetries of an Ising model,
should it be a magnet, an alloy or anything else, will exhibit a
diverging susceptibility $\chi\sim|T-T_c|^{-7/4}$ (see a sketch in
Fig.~\ref{fig:ChiMain}) with the precise value $7/4$ for the
exponent.

On a theoretical ground, the major two-dimensional problems (Ising
model and its generalizations like the Potts model and the
percolation model, $XY$ model, Heisenberg model and so on) are
essentially solved at least for their critical singularities (when a
second-order phase transition is indeed present), but critical
exponents are not the only universal quantities at a critical point.
The universal character of appropriate combinations of critical
amplitudes~\cite{PrivmanHohenbergAharony91} is also an important
prediction of scaling theory but  in some cases these combinations
remain uncompletely determined and subject to controversies.

The Potts model~\cite{Potts52,Wu82}, as one of the paradigmatic
models exhibiting continuous phase transitions is a good frame to
consider the question of universal combinations of amplitudes. The
universality class of the Potts model at its critical point is
parametrized by the number of states $q$. The two-dimensional Potts
model with three and four states can be experimentally realized as
strongly chemisorbed atomic adsorbates on metallic surfaces at
sub-monolayer concentrations~\cite{SokolowskiPfnuer94}. Although
critical exponents could be measured quite accurately for adsorbed
sub-monolayers, confirming that these systems actually belong to the
three-state~\cite{NakajimaEtAl97} or to the four-state Potts model
classes~\cite{VogesPfnuer98}, it is unlikely that the low
temperature LEED results can be pushed~\cite{PfnuerPrivateComm} to
determine also the critical amplitudes. Therefore, the numerical
analysis of these models is the only available tool to check
analytic predictions.

The critical amplitudes and critical exponents describe the behaviour of
the magnetization $m$, the susceptibility $\chi$, the specific heat $C$ and
the correlation length $\xi$ for a spin system in zero external
field\footnote{In this paper we only deal with the physical properties
in zero magnetic field.} in
the vicinity of the critical point
\begin{eqnarray}
    M(\tau) &\approx& B_- (-\tau)^\beta,\ \tau <0, \label{m-crit}\\
    \chi(\tau) &\approx& \Gamma_\pm \tmod^{-\gamma}, \label{x-crit}\\
    \chi_T(\tau)&\approx& \Gamma_T (-\tau)^{-\gamma},\ \tau <0,\label{xt-crit}\\
    C(\tau) &\approx& \frac{A_\pm}{\alpha}\tmod^{-\alpha},
    \label{c-crit}\\
    \xi(\tau) &\approx& \xi_0^\pm \tmod^{-\nu}.
    \label{k-crit} \end{eqnarray}
 Here $\tau$ is the reduced temperature $\tau=(T-T_c)/T$ and the
labels $\pm$ refer to the high-temperature and low-temperature sides
of the critical temperature $T_c$. For the Potts models
 with $q>2$ a transverse susceptibility $\chi_T$ can be defined in the
low-temperature phase\footnote{In
the following we will use equally the notations $\Gamma_L$ or $\Gamma_-$ for
the longitudinal susceptibility amplitude in the low
temperature phase.}.

Critical exponents are known exactly for 2D Potts
model~\cite{denNijs79,Pearson80,Nienhuis84,DotsenkoFateev84} through the
relation $x_\epsilon=(1-\alpha)/\nu$ to the thermal scaling dimension
\begin{equation}
    x_\epsilon=\frac{1+y}{2-y}
    \label{x-T}
\end{equation}
and the relation $x_\sigma=\beta/\nu$ to the magnetic scaling
dimension
\begin{equation}
    x_\sigma=\frac{1-y^2}{4(2-y)},
    \label{x-m}
\end{equation}
where the parameter $y$ is related to the number of states $q$
of the Potts variable by the expression
\begin{equation}
    \cos\frac{\pi y}{2}=\frac12 \sqrt{q}
    \label{y-from-q}
\end{equation}
The central charge of the corresponding conformal
field
theory is also simply expressed~\cite{DotsenkoFateev84} in terms of $y$
\begin{equation}
    c=1-\frac{3y^2}{2-y}.
    \label{central-c}
\end{equation}

Analytical estimates of  critical amplitude ratios for
the $q$-state Potts models with $q=1$, $2$, $3$, and $4$ were
 recently obtained
by Delfino and Cardy~\cite{DelfinoCardy98}. They used the
two-dimensional scattering field theory of Chim and
Zamolodchikov~\cite{ChimZamolodchikov92} and estimated the central
charge $c=0.985$ for 4-state Potts model, for which the exactly known
value is $c=1$.  Reporting
these approximate values in (\ref{central-c}), one can calculate
the scaling dimensions from
(\ref{x-T})-(\ref{x-m}) and get
the values $x_\sigma=0.13016$ and
$x_\epsilon=0.577$, to be compared
respectively to the exact values $1/8$ and
$1/2$. The discrepancy  is around
4 and 15 per cent, emphasizing the difficulty
of the  $q=4$ case (to give an idea, in the case of the 3-state
Potts model, a similar analysis leads to a very good agreement with
less than one percent deviation).

The universal susceptibility amplitude ratios
$\Gamma_+/\Gamma_L$ and $\Gamma_T/\Gamma_L$
were also calculated
in~\cite{DelfinoCardy98} and \cite{DelfinoBarkemaCardy00}.
The figures obtained are the following,
\begin{eqnarray}
q=3:&\Gamma_+/\Gamma_L=13.848, & \Gamma_T/\Gamma_L=0.327 ,
    \label{res-theor-q3}\\
q=4:&\Gamma_+/\Gamma_L=4.013, & \Gamma_T/\Gamma_L=0.129 .
    \label{res-theor-q4}
\end{eqnarray}
These results have been confirmed numerically in the case $q=3$ by
several groups, $\Gamma_+/\Gamma_L\approx 10$ and
$\Gamma_T/\Gamma_L\approx 0.333(7)$ in
Ref.~\cite{DelfinoBarkemaCardy00} (Monte Carlo (MC) simulations),
$\Gamma_+/\Gamma_L=14\pm 1$ in Ref.~\cite{ShchurButeraBerche02} (MC
and series expansion (SE) data) and quite recently, these results
were confirmed and substantially improved,
$\Gamma_+/\Gamma_L=13.83(9)$, $\Gamma_T/\Gamma_L=0.325(2)$, by
Enting and Guttmann~\cite{EntingGuttmann03} who analysed new longer
series expansions.

The 4-state Potts model was also studied through MC simulations in
Ref.~\cite{DelfinoBarkemaCardy00}, but the authors considered that
their data were not conclusive. Another MC contribution is reported
by Caselle, et al~\cite{CaselleTateoVinci99},
$\Gamma_+/\Gamma_L=3.14(70)$, and Enting and
Guttmann~\cite{EntingGuttmann03} also analysed SE data for the
4-state Potts model and found $\Gamma_+/\Gamma_L=3.5(4)$,
$\Gamma_T/\Gamma_L=0.11(4)$ in relatively good agreement with the
predictions of~\cite{DelfinoCardy98}
and~\cite{DelfinoBarkemaCardy00}. The situation thus seems to be
clear, although the use of the logarithmic corrections in the
fitting procedure of MC data was questioned, e.g. in
\cite{EntingGuttmann03}: {\em [Caselle et al] estimates depend
critically on the assumed form of the sub-dominant terms, and on the
further assumption that the other sub-dominant terms, which include
powers of logarithms, powers of logarithms of logarithms etc, can
all be neglected. We doubt that this is true.}

Let us recall that the existence of  logarithmic corrections to
scaling in the 4-state Potts model was pointed out in  the
pioneering works of Cardy, Nauenberg and
Scalapino~\cite{CardyNauenbergScalapino80,NauenbergScalapino80},
where a set of non-linear RG equations were proposed. Their
discussion was later extended by Salas and
Sokal~\cite{SalasSokal97}.

Generically, the logarithmic corrections appear as corrections to
scaling. We mentioned above that in the vicinity of a critical
point, a susceptibility for example diverges like
$\chi(\tau)\approx\Gamma_\pm\tmod^{-\gamma}$. This is true, but this
singular behaviour can be superimposed to a regular signal (e.g.
$D_0+D_1\tmod+\dots$), and the leading singular behaviour itself
needs to be corrected when we consider the physical quantity away
from the transition temperature. The expression for $\chi(\tau)$
then takes a form which can become ``terrific'':

\vspace{2mm}
\begin{tabular}{lll}
$\chi(\tau) =$ & $D_0+D_1\tmod+\dots$ & regular background \\
\noalign{\vskip2pt}
   \ \hfil$+$  & $\Gamma\tmod^{-\gamma}(1 +
    $ & leading singularity \\
\noalign{\vskip2pt}
& $\qquad\qquad
+a^{(1)}\tmod^{\Delta}
            +a^{(2)}\tmod^{2\Delta}+\dots$
            & leading corrections  \\
\noalign{\vskip2pt}
           & $\qquad\qquad+{a'}^{(1)}\tmod^{\Delta'}
        +{a'}^{(2)}\tmod^{2\Delta'}+\dots$
        & next corrections \\
\noalign{\vskip2pt}
           & $\qquad\qquad+b^{(1)}\tmod
        +b^{(2)}\tmod^{2}+\dots)$
        & analytic corrections \\
\noalign{\vskip2pt}
               & $\times(-\ln\tmod)^\star\times
    \left(1+\triangleleft\frac{\ln(-\ln\tmod)}{-\ln\tmod}\right)
    \times \dots$
        & logarithmic corrections
\end{tabular}

\vspace{2mm} \noindent Together with the amplitude and exponent
associated to the leading singularity, $\Gamma$ and $\gamma$, appear
corrections to scaling due to the presence of irrelevant scaling
fields ($a^{(n)}$ and $\Delta$, ${a'}^{(n)}$ and $\Delta'$, \dots),
analytic corrections due to non-linearities of the relevant scaling
fields ($b^{(n)}$), or multiplicative  logarithmic corrections
($\triangleleft$ and $\star$,  \dots). These logarithmic
coefficients may have different origins (see e.g. in
Ref.~\cite{PrivmanHohenbergAharony91} and references therein). They
can be due to the upper critical dimension, to poles in the
expansion of regular and singular amplitudes, or to the presence of
marginal scaling fields. The 4-state Potts model belongs to this
latter category.

Some of the quantities indicated above are universal. This is the
case of the exponents as well as of many combinations of
coefficients. In the present paper we are interested in the
amplitude of the leading singular term, but its precise
determination can be affected by the form of the corrections. We
shall be concerned with the following universal combinations
 of critical amplitudes
\begin{equation}
    \frac{A_+}{A_-}, \;\; \frac{\Gamma_+}{\Gamma_L}, \;\;
    \frac{\Gamma_T}{\Gamma_L}, \;\; R_C^+=\frac{A_+\Gamma_+}{B_-^2}.
    \label{univ-rat}
\end{equation}
We present, for the 4-state Potts model, more accurate Monte Carlo data
supplemented by a reanalysis of the extended series made available
by Enting and Guttmann~\cite{EntingGuttmann03} and
we address the following question:
Is it possible to devise some procedure in which the
role of these logarithmic corrections is properly taken into account?

\section{Amplitudes and universal combinations}
\label{sec:2}  
The scaling hypothesis states that the singular part of the
free energy density can be written in terms of the deviation from the
critical point, $\tau=(T-T_c)/T$ and $h=H-H_c$,
\be
f_{\rm sing}(\tau,h)=b^{-D}F_\pm(\kappa_\tau b^\yTau\tau,\kappa_h b^\yH h)
\label{eq-fsing}\ee
where $F_\pm(x,y)$ is a universal function (actually there is one
universal function for each side $\tau>0$ or $\tau<0$ of the critical point)
and $\kappa_\tau$ and $\kappa_h$ are ``metric factors'' which contain
all the non universal aspects of the critical behaviour.
$D$ is the space dimension.
 Let us stress that
the functions $F_\pm$ are universal in the sense that some details of the
model are irrelevant (e.g. the coordination number
of the lattice (so long as it
remains finite), the presence of next nearest neighbour interactions, etc)
but they depend on the boundary conditions or the shape of the
system. The metric factors on the other hand depend on these details, and
the universal combinations are obtained when the metric factors are eliminated
from some combinations.

The connection with scaling relations can be shown with an example.
From Eq.~(\ref{eq-fsing}), we also deduce similar homogeneous expressions
for the magnetization, $M(\tau,h)=b^{-D+\yH}\kappa_h M_\pm(x,y)$ and the
susceptibility, $\chi(\tau,h)=b^{-D+2\yH}\kappa_h^2 X_\pm(x,y)$. The choice
$b=(\kappa_\tau\tmod)^{-1/\yTau}$ and $h=0$ leads (for example below the
transition temperature) for the following combination
of quantities
\be
    \alpha\frac{C(\tau,0)\chi(\tau,0)}{m^2(\tau,0)}\tmod^{2}\equiv
    (\kappa_\tau\tmod)^{2-\alpha-2\beta-\gamma}\alpha
    \frac{C_-(1,0)X_-(1,0)}{M_-^2(1,0)}\equiv R.
\ee The prefactor takes the value 1 thanks to the well known scaling
relation between critical exponents $\alpha+2\beta+\gamma=2$. Thus
it follows that the above combination is a {\em universal number}.
From the definition of magnetization, specific heat and
susceptibility amplitudes in zero magnetic field, e.g.
$M(\tau,0)=\tmod^{\frac{D-\yH}{\yTau}}\kappa_h M_-(1,0)\equiv
B_-\tmod^\beta$ by virtue of Eq.~(\ref{m-crit}), this universal
number is in fact a combination of amplitudes, $R\equiv
A_-\Gamma_-/B_-^2$. We have similar universal combinations above the
critical temperature or associated to other scaling relations.

\section{RG approach for the Potts model and logarithmic corrections at $q=4$}
\label{sec:3}  
\begin{figure}[ht]
\begin{center}
\includegraphics[angle=0,scale=0.40]{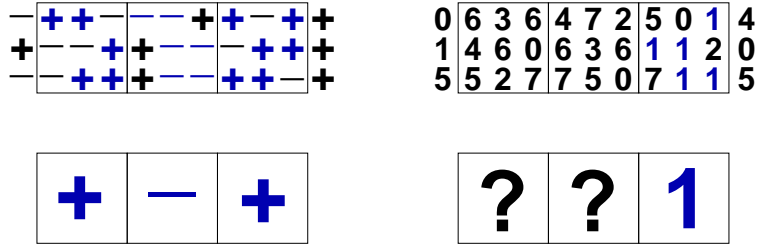}
\end{center}\vspace*{-0.4cm}
\caption{\label{fig:PottsMain}
Decimation of spin blocks for the Ising model (left) and high-$q$ Potts
model (right). In the latter case, the state of many cells cannot be decided
by a simple majority rule.}
\end{figure}

Let us remind that the $q$-state Potts model is an extension of the
usual lattice Ising model in which the site variables $s_i$
(abusively called spins) can have $q$ different values,
$s_i=0,1,\dots\  q-1$ but the nearest neighbour interaction energy
$-J\delta_{s_i,s_j}$ only takes two possible values, e.g. $-J$ and
$0$ depending whether the neighbouring spins are in the same state
or not.  The Hamiltonian of the model reads as
\begin{equation}
    H = - J\sum_{\langle ij \rangle}\delta_{s_i s_j}\; .
    \label{Ham}
\end{equation}
At the early times of real-space renormalization, the application to the pure
Potts model led to some difficulties:  the impossibility to affect
a particular value for the spin of a cell after decimation due to a too
large number of states (see Fig.~\ref{fig:PottsMain}).

It was thus necessary to extend the parameter space, introducing new
variables, called vacancies, to replace the question marks in
Fig.~(\ref{fig:PottsMain}) and to study an annealed disordered
model. The RG equations satisfied by the model, written in terms of
the relevant thermal and magnetic fields $\thermalfield$ and $h$,
with corresponding RG eigenvalues $\yT $ and $\yH $, and the
marginal dilution field $\psi$, are given by
%
 $   \frac{d\thermalfield }{d \ln b}=\yT\thermalfield$,
 $   \frac{dh }{d \ln b}=\yH h$,
 $   \frac{d\psi }{d \ln b}=q-q_c$,
%
where $b$ is the length rescaling factor and $l=\ln b$. When $q> q_c$,
the dilution field $\psi$ is relevant (and the phase transition is of
first order), while in the regime $q<q_c$, $\psi$ is irrelevant and the
system exhibits a second-order phase transition.
The case $q=q_c$ is marginal.
This picture is qualitatively correct, and in fact the critical value of
the number of states which discriminates between the two regimes is
$q_c=4$. In the $q$ direction, $q_c=4$ appears as the end of a line
of fixed points where logarithmic corrections are expected.
At $q_c=4$, the RG equations 
were extended by Cardy, Nauenberg and Scalapino
(CNS)~\cite{NauenbergScalapino80,CardyNauenbergScalapino80} and then
by Salas and Sokal (SS)~\cite{SalasSokal97}. As a result of the
coupling between the dilution field $\psi$,  and $\tau$ and $h$,
they were led to non-linear equations,
\begin{eqnarray}
    \frac{d\thermalfield }{d \ln b}&=&(\yT +\yTpsi \psi )\thermalfield
        ,\label{bb-eq4}\\
    \frac{dh }{d \ln b}&=&(\yH +\yHpsi \psi)h,
        \label{bb-eq5}\\
    \frac{d\psi }{d \ln b}&=&g(\psi).
        \label{bb-eq6}
\end{eqnarray}
The function $g(\psi)$ may be Taylor expanded,
$g(\psi)=\ypsideux \psi^2(1+\frac{\ypsitrois }{\ypsideux }\psi+\dots)$.
Accounting for marginality of the dilution field, there is no
linear term at $q=4$.
Comparing to the available results (for example the expression
of the latent heat for $q\ge q_c$ by Baxter~\cite{Baxter73} or the
den Nijs and Pearson's conjectures for the RG eigenvalues for
$q\le q_c$~\cite{denNijs79,Pearson80}),
the parameters were found to
take the values  $\yTpsi =3/(4\pi)$,
$\yHpsi =1/(16\pi)$, $\ypsideux =1/\pi$ and
$\ypsitrois =-1/(2\pi^2)$, while
the relevant scaling dimensions are
$\yT =\nu^{-1}=3/2$ and $\yH =15/8$.

The fixed point is at $\thermalfield=h=0$. Starting from initial conditions
$\thermalfieldzero$, $\hzero$, the relevant fields grow exponentially with
$l$  up to some $\thermalfieldl=O(1)$, $h=O(1)$ outside the
critical region. Notice also that the marginal field $\psil$
remains of order of its initial value, $\psi\sim O(\psizero)$.
In zero magnetic field, under a change of length
scale, the singular part of the free energy density transforms
according to \be
    f(\psizero,\thermalfieldzero)=e^{-Dl}f(\psil,1).\label{ap-eq4}
\ee
Solving \Eqs{bb-eq4}{bb-eq6} leads to
\be
    l=-\frac 1{\yT }\ln \thermalfieldzero
    +\frac{\yTpsi }{\yT \ypsideux }
    \ln \left(\frac {\psizero}{\psil} G(\psizero,\psil)\right)
,
    \label{ap-eq5}
\ee
(for brevity we will denote
$\nu=1/\yT=\frac 23$, $\mu=\frac{\yTpsi }{\yT\ypsideux }=\frac 12$).
Note that $G(\psizero,\psil)$ would take the value 1 in
Ref.~\cite{CardyNauenbergScalapino80}
and the value $\frac{\ypsideux +\ypsitrois \psil}
{\ypsideux +\ypsitrois \psizero}$ in Ref.~\cite{SalasSokal97}.
We can thus deduce the following behaviour for the free energy density
in zero magnetic field in
terms of the thermal and dilution fields,
\be
    f(\thermalfieldzero,\psizero)=
    \thermalfieldzero^{D\nu}
     \left(
    \frac {\psil}{\psizero}
    \frac{\ypsideux +\ypsitrois \psizero}
    {\ypsideux +\ypsitrois \psil}
     \right)^{D\mu}
    f(1,\psil).
    \label{ap-eq6}
\ee A similar expression would be obtained if the magnetic field $h$
were also included. The other thermodynamic properties follow from
derivatives with respect to the scaling fields. The quantity between
parentheses is the only one where the log terms are hidden in the
4-state Potts model, and thus we may infer that {\em not only the
leading log terms}, but {\em all the log terms hidden in the
dependence on the marginal dilution field} disappear in the
conveniently defined effective ratios\footnote{i.e. effective ratios
which eventually tend towards universal limits when $\tmod\to 0$}.
Now we proceed by iterations of \Eq{ap-eq5}, and eventually we get
for the full correction to scaling variable the {\em heavy}
expression
\begin{eqnarray}
    \frac {\psizero}{\psil} G(\psizero,\psil)&=&
    {\rm const}\times
    \underbrace{(-\ln|\tau|)}_{\hbox{CNS}}
\underbrace{    \left(
        1+\frac 34
        \frac{\ln(-\ln|\tau|)}
            {-\ln|\tau|}\right)
    \left(
    1-\frac 34\frac{\ln(-\ln|\tau|)}{-\ln|\tau|}\right)^{-1}
    }_{1+\frac 32
        \frac{\ln(-\ln|\tau|)}
            {-\ln|\tau|} \ \hbox{in SS}}\nonumber \\
& &  \quad  \times
    \left(
    1+\frac 34\frac{1}{(-\ln|\tau|)}
    \right)\times\underbrace{
\left(1+\frac{{\rm const}}
{-\ln|\tau|}
+O\left(\frac{1}{-\ln|\tau|^2}\right)\right)
}_{{F}(-\ln|\tau|)}
\label{eq-heavy}
\end{eqnarray}
where CNS and SS refer to the results previously obtained in the
literature~\cite{NauenbergScalapino80,CardyNauenbergScalapino80,SalasSokal97}
and
${F}(-\ln|\tau|)$ is the only factor where
{\em non universality} enters through
the dilution field $\psizero$.
This allows to write down the behaviour of the magnetization for example
\begin{eqnarray}
    M(\tau)&=&B_-|\tau|^{1/12}(-\ln|\tau|)^{-1/8}
    \left[
    \left(1+\frac 34\frac{\ln(-\ln|\tau|)}{-\ln|\tau|}\right)
        \right.\nonumber\\
    &&\left.
    \ \qquad
    \left(1-\frac 34\frac{\ln(-\ln|\tau|)}{-\ln|\tau|}\right)^{-1}
    \left(1+\frac 34\frac{1}{-\ln|\tau|}\right)
    {F}(-\ln|\tau|)
        \right]^{-1/8}.
\label{eq-Mfinal}
\end{eqnarray}

\section{Numerical techniques}
\label{sec-MCSE} In the Monte Carlo simulations we use the Wolff
algorithm~\cite{Wolff89} for studying square lattices of linear size
$L$ (between $L=20$ and $L=200$) with periodic boundary conditions.
Starting from an ordered state, we let the system equilibrate in
$10^5$ steps measured by the number of flipped Wolff clusters. The
averages are computed over $10^6$---$10^7$ steps. The random numbers
are produced by an exclusive-XOR combination of two shift-register
generators with the taps (9689,471) and (4423,1393), which are
known~\cite{Shchur99} to be safe for the Wolff algorithm.

The order parameter of a microstate
${\tt M}({\tt t})$ is evaluated during the simulations as
\begin{equation}
    {\tt M}=\frac{qN_m/N-1}{q-1},
    \label{Order-Potts}
\end{equation}
where $N_m$ is the number of sites $i$ with $s_i=m$ at the time $\tt
t$ of the simulation and $m\in [0,1,...,(q-1)]$ is the spin value of
the majority of the sites. $N=L^2$ is the total number of spins. The
thermal average is denoted $M=\<{\tt M}\>$. Thus, the longitudinal
susceptibility in the low-temperature phase is measured by the
fluctuation of the majority of the spins
\begin{equation}
    \chi_L={\beta} (\langle N_m^2\rangle-\langle N_m\rangle^2)
    \label{susc-LT}
\end{equation}
and the transverse susceptibility is defined in the low-temperature
phase as the fluctuations of the minority of the spins
\begin{equation}
    \chi_T=\frac{\beta}{(q-1)}\sum_{\mu\ne m}
    (\langle N_\mu^2\rangle-\langle N_\mu\rangle^2),
    \label{susc-T}
\end{equation}
while in the high-temperature phase $\chi_+$ is given by the fluctuations
in all $q$ states,
\begin{equation}
    \chi_+=\frac{\beta}{q}\sum_{\mu=0}^{q-1}
    (\langle N_\mu^2\rangle - \langle N_\mu\rangle^2),
    \label{susc-HT}
\end{equation}
where $N_\mu$ is the number of sites with the spin in the state $\mu$.
The internal energy density of a microstate is calculated as
\begin{equation}
    {\tt E}=-\frac{1}{N} \sum_{\langle ij \rangle}\delta_{s_i s_j}\,
    \label{energy}
\end{equation}
its ensemble average denoted as $E=\<{\tt E}\>$
and the specific heat per spin measures the energy fluctuations,
\begin{equation}
    C=-\beta^2\frac{\partial  E }{\partial\beta}
    =\beta^2\left(\langle {\tt E}^2 \rangle - \langle {\tt E} \rangle^2 \right).
    \label{heat}
\end{equation}

Our MC study of the critical amplitudes is supplemented by an
analysis of the high-temperature (HT) and low-temperature (LT)
expansions for $q=4$ recently
calculated through remarkably high orders
by Enting, Guttmann and
coworkers~\cite{BriggsEntingGuttmann94,EntingGuttmann03}. In terms
of these series, we can compute the effective critical amplitudes for
the susceptibilities  and the magnetization and
extrapolate them by the current resummation techniques, namely simple
Pad\'e approximants (PA)  and differential approximants (DA)
properly biased with the exactly known critical temperatures and
critical exponents.
The LT expansions, expressed in terms of the variable $z=
\exp(-\beta)$, extend through  $z^{59}$
for the
longitudinal susceptibility and
through $z^{47}$ in the case of the transverse
susceptibility. The
magnetization and energy expansions extend through $z^{43}$.
The HT expansion is computed in terms of the variable
$v=(1-z)/(1+(q-1)z)$. The susceptibility expansion
has been computed up to  $v^{24}$ and the energy
expansion up to $v^{43}$.

\section{Analysis of the magnetization behaviour}
\label{sec:5}  
For the sake of simplification of the notations,
we group all the terms containing logs in \Eq{eq-heavy}
into a single function
${H}(-\ln|\tau|)={E}(-\ln|\tau|)\times {F}(-\ln|\tau|)$ where
\bey
    {E}(-\ln|\tau|)&=&
    \left[
    (-\ln|\tau|)
    \left(1+\frac 34\frac{\ln(-\ln|\tau|)}{-\ln|\tau|}\right)
        \right.\nonumber\\
    &&\left.
    \qquad\times \left(1-\frac 34\frac{\ln(-\ln|\tau|)}{-\ln|\tau|}\right)^{-1}
    \left(1+\frac 34\frac{1}{-\ln|\tau|}\right)
        \right].\label{ap-eq99}
\eey
The function $E$ contains all leading logarithms with universal coefficients,
it
is known exactly while the function $F$ needs to be fitted.
We thus obtain a closed
expression for the dominant logarithmic corrections which
is more suitable than previously proposed forms
to describe an observable (Obs.) in
the temperature range accessible in a numerical study:
\bey
    {\rm Obs.}(\tau)
        &\simeq&{\rm Ampl.}\times
    \tmod^{\bullet}\times
    H^{\sharp}(-\ln\tmod)\times(1+{\rm Corr.\ terms})
,
        \label{eq-Obs_us}\\
    {\rm Corr.\ terms}&=&a \tmod^{2/3}+b_\pm\tmod+\dots,
\label{eq-Obs} \eey where ${\bullet}$ and $\sharp$ are exponents
which depend on the observable considered, and take the values
$1/12$ and $-1/8$ respectively in the case of the magnetization.
Here we stress that the inclusion of a correction in $|\tau|^{2/3} $
seems to be necessary according to previous work of Joyce on the
Baxter-Wu model~\cite{Joyce75a,Joyce75b} (of 4-state Potts model
universality class), where the magnetization is shown to obey an
expression of the form $ M(\tau)=B_-\tmod^{1/12}(1+{\rm
const}\times\tmod^{2/3}
    +{\rm const}'\times\tmod^{4/3})$.
The exponent $2/3$  comes out from the conformal scaling dimensions
of Dotsenko and Fateev~\cite{DotsenkoFateev84}, and its presence is
needed in order to account for the numerical results (see also
Ref.~\cite{ShchurBercheButera07}). Caselle et
al.~\cite{CaselleTateoVinci99} also considered $|\tau|^{2/3} $ term
to fit the magnetization. Here we also allow inclusion of a linear
correction in $b_\pm\tmod$ to account for possible non linearities
of the relevant scaling fields~\cite{PrivmanHohenbergAharony91}. The
next term in $\tmod^{4/3}$ will be forgotten.

\begin{figure}[ht]
\begin{center}
\includegraphics[angle=0,scale=0.40]{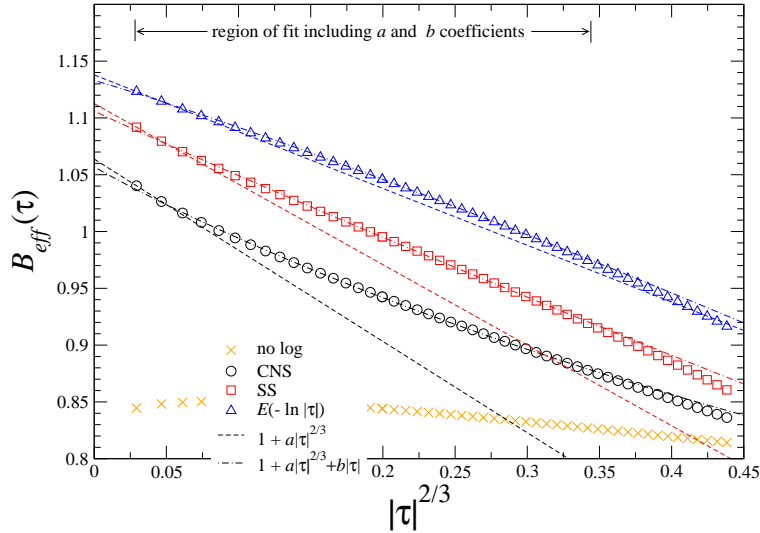}
\end{center}\vspace*{-0.4cm}
\caption{\label{fig:Fit_M}
Effective amplitudes (from MC data) as deduced from different assumptions
for the logarithmic corrections. Symbols correspond to MC data,
dashed and dotted-dashed lines are fits as explained in the text.}
\end{figure}

In Fig.~\ref{fig:Fit_M}
we plot  effective magnetization amplitudes $B_{eff}(\tau)$
vs $\tmod^{2/3}$.
From the available data for the magnetization,
we define the following quantities,
\bey
B_{CNS}(\tau)&=&M(\tau)\tmod^{-1/12}(-\ln\tmod)^{1/8},\label{BCNS}\\
B_{SS}(\tau)& =&M(\tau)\tmod^{-1/12}(-\ln\tmod)^{1/8}\left(
    1-\frac 3{16}\frac{\ln(-\ln\tmod)}{-\ln\tmod}
    \right)^{-1},\label{BSS}\\
B_{E}(\tau)&  =&M(\tau)\tmod^{-1/12}E^{1/8}(-\ln\tmod),\label{BE}
\eey which are expected to behave according to the corrections to
scaling \be  B_-(1+a \tmod^{2/3}+b_\pm\tmod+\dots)\label{Btry}\ee
The numerical results are shown in Fig.~\ref{fig:Fit_M}. From bottom
to top the various symbols indicate the effective amplitudes
 with "no-log" at all,  then with the CNS and the SS corrections and finally
the effective amplitude where the known universal logarithmic terms
 have been included.
 The dashed lines correspond to a rough
determination of the corrections to scaling including only the terms
in $a\tmod^{2/3}$ in the limit $\tmod\to 0$, and the dot-dashed
lines include also the terms in $b\tmod$. From this plot, we deduce
that {\em none of the three effective amplitudes in \Eqs{BCNS}{BE}
can be correctly fitted by \Eq{Btry}}, since the coefficients of the
correction terms (e.g. the coefficient $a$ which is estimated
directly by the slope at small $\tmod$ values) strongly depend on
the range of fit. This undesirable dependence of
 the coefficients on the width of the temperature window is shown
 in the first six lines of table~\ref{tab1}.

\begin{table}[h]
\caption{Fits of the effective amplitude of the magnetization. }
\center
\begin{tabular}{lllll}
\hline\noalign{\vskip-1pt}\hline\noalign{\vskip2pt}
$B_{eff}(\tau)$\phantom{---} & $\tmod^{2/3}$-window\phantom{---}
    & \phantom{-}$B_-$\phantom{---} & \phantom{---}$a$\phantom{---}
    & \phantom{---}$b$\phantom{---} \\
 \noalign{\vskip2pt}\hline
$B_{CNS}(\tau)$\  & $[0,0.15]$ & $1.07$ & $-0.98$ & $\phantom-0.94$ \\
                  & $[0,0.45]$ & $1.05$ & $-0.66$ & $\phantom-0.29$ \\
 \noalign{\vskip2pt}\hline
$B_{SS}(\tau)$\   & $[0,0.15]$ & $1.11$ & $-0.77$ & $\phantom-0.56$ \\
                  & $[0,0.45]$ & $1.10$ & $-0.45$ & $-0.06$ \\
 \noalign{\vskip2pt}\hline
$B_{E}(\tau)$\    & $[0,0.15]$ & $1.14$ & $-0.47$ & $\phantom-0.16$ \\
                  & $[0,0.45]$ & $1.13$ & $-0.25$ & $-0.27$ \\
 \noalign{\vskip2pt}\hline
$B_{EF}(\tau)$\   & $[0,0.15]$ & $1.16$ & $-0.20$ & $\phantom-0.02$ \\
                  & $[0,0.45]$ & $1.16$ & $-0.18$ & $-0.02$ \\
\hline\noalign{\vskip-1pt}\hline\noalign{\vskip2pt}
\end{tabular}
\label{tab1}
\end{table}

\begin{figure}[ht]
\begin{center}
\includegraphics[angle=0,scale=0.40]{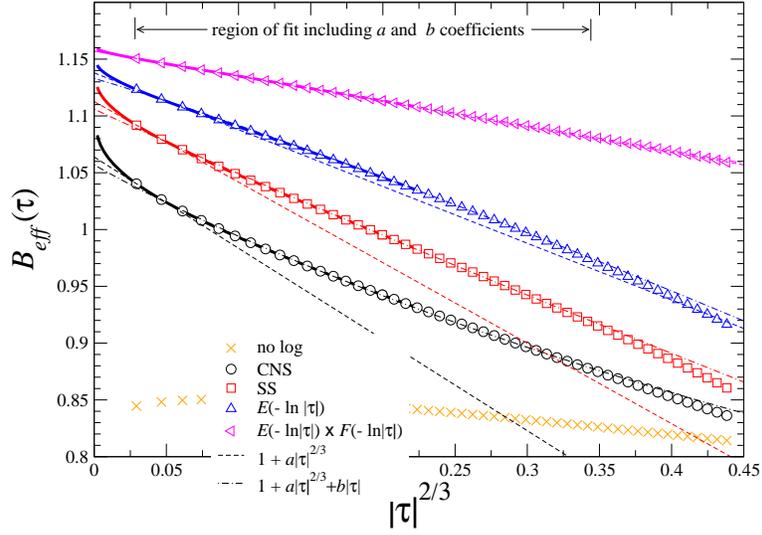}
\end{center}\vspace*{-0.4cm}
\caption{\label{fig:Fit_MSE}
Effective amplitudes (from MC and SE data) as deduced from
different assumptions
for the logarithmic corrections. The upper curves correspond to
$B_{EF}(\tau)$ and the thick solid lines to SE data.}
\end{figure}

In order to improve the quality of the fits, one has to take into
account the correction function $F(-\ln\tmod)$ and to extract an
effective function ${F}_{eff}(-\ln|\tau|)$ which mimics the real one
in the convenient temperature range. This is done by fitting
$B_{eff}(\tau)$ to a more complicated expression, \be  B_-(1+a
\tmod^{2/3}+b_\pm\tmod+\dots)\times \left(1+\frac{C_1}{-\ln|\tau|}
+\frac{C_2\ln(-\ln|\tau|)}{(-\ln|\tau|)^2}\right)^{1/8},\label{Btry2}\ee
which means that we include the corrections to scaling {\em and} the
non universal function function $F(-\ln|\tau|)$ taking the
approximate expression \be {F}_{eff}(-\ln|\tau|)\simeq
\left(1+\frac{C_1}{-\ln|\tau|}
+\frac{C_2\ln(-\ln|\tau|)}{(-\ln|\tau|)^2}\right)^{-1}.
\label{eq-F_function}\ee While $a$ and $b$  are coefficients of
corrections to scaling due to irrelevant operators, $C_1$ and $C_2$
are {\em effective} coefficients of logarithmic terms which, in a
given temperature range, mimic a slowly convergent series of
logarithmic terms depending on a non universal dilution field.
Therefore, we expect that different fits made in different
temperature windows will produce different values of $C_1$ and $C_2$
while $a$ and $b$ (and of course also the magnetization amplitude
$B_-$) should be relatively less influenced by the window range. The
choice of values for $C_1$ and $C_2$ is thus partially arbitrary and
the values quoted should be specified together with the temperature
window where they are appropriate. In the following, we obtain
$C_1\simeq-0.76$ and $C_2\simeq-0.52$ in the window $\tmod^{2/3}\in
[0,0.35]$, which yields an amplitude $B_-\simeq 1.157$. The
resulting $a$ and $b$ coefficients now appear very stable. This is
checked in  Fig.~\ref{fig:Fit_MSE} where the quantity \be
B_{EF}(\tau) =M(\tau)\tmod^{-1/12}[E(-\ln\tmod)F(-\ln\tmod)]^{1/8}
\ee is reported toghether with the previous curves and fitted as
indicated in table~\ref{tab1}. As an independent test, we add the SE
data which are superimposed to the MC data at small values of
$\tmod$ only for this latter assumption of effective amplitude.

\begin{figure}[ht]
\begin{center}
\includegraphics[angle=0,scale=0.40]{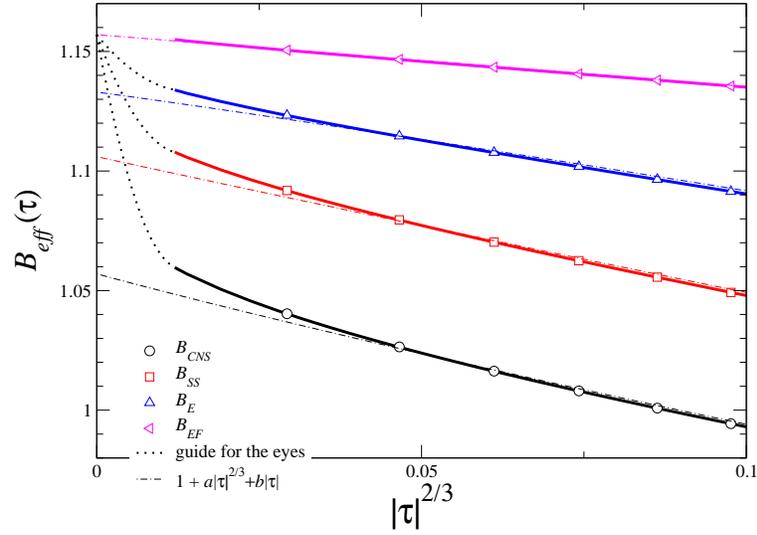}
\end{center}\vspace*{-0.4cm}
\caption{\label{fig:Fit_M_Zoom}
Zoom of the effective amplitudes in the vicinity of the critical point.
MC data (symbols) and SE data (thick solid lines).}
\end{figure}

Eventually, our approach confirms expression
\Eq{eq-Mfinal} for the magnetization, with the function $F(-\ln\tmod)$
given in \Eq{eq-F_function} and the parameters $C_1$ and $C_2$ given above for
the appropriate temperature window. Nevertheless, we have to stress that
the different effective amplitudes should all reach {\em the same amplitude
$B_-$ in the limit $\tmod\to 0$}, since $B_{CNS}(\tau)$ is in fact
an approximation of $B_{SS}(\tau)$, which is an approximation of
$B_{E}(\tau)$, which eventually approximates $B_{EF}(\tau)$. An
attempt of illustration of this behaviour is shown in
Fig.~\ref{fig:Fit_M_Zoom} where dotted lines (which are only guides  for the
eyes) all converge towards the unique value $B_-\simeq 1.157$. Here we stress
that we have deleted the points of SE data which are too close to the
critical point, since the series are no longer reliable
because the current extrapolation procedures are in principle
unable to approximate the complicated structure of the singularity
involving log corrections.

\section{Universal combinations for the 4-state Potts model and conclusions}
\label{sec:6}  
The other quantities can be analyzed along the same lines. The
important point is that now the function $F_{eff}(\tau)$ is fixed in
the corresponding temperature scale and thus the remaining freedom
for the other physical quantities in \Eq{eq-Obs_us} is only through
the leading amplitude and the coefficients of $\tmod^{2/3}$ and
$\tmod$-terms in the corrections plus possibly the background terms.
It is still a complicated task to perform this analysis, but the
current results for the universal combinations mentioned in the
introduction appear in the following table.

\begin{table}[h]
\caption{Rough estimate of the universal
combinations of the critical amplitudes in the 4-state
    Potts model. }
\center
\begin{tabular}{llllc}
\hline\noalign{\vskip-1pt}\hline\noalign{\vskip2pt}
$A_+/A_-$\qquad\qquad & $\Gamma_+/\Gamma_L$\qquad\qquad
    & $\Gamma_T/\Gamma_L$\qquad\qquad
    & $R_C^+$\qquad\qquad & source
\\ \noalign{\vskip2pt}\hline
   $1.^{a}$       &  $4.013$            & $0.129$             & $0.0204$      &
     \cite{DelfinoCardy98,DelfinoBarkemaCardy00}\\
          &  $3.14(70)$              &                     &   $0.021(5)$ &
     \cite{CaselleTateoVinci99}\\
          &  $3.5(4)$   & $0.11(4)$                     &       &
     \cite{EntingGuttmann03}\\
$1.00(1)$ & $6.7(4)$ & $0.161(3)$ &$0.0307(2)$ &
     {here}\\
\hline\noalign{\vskip-1pt}\hline\noalign{\vskip2pt}
\multicolumn{5}{l}{$^a$  exact result from duality}
\end{tabular}
\label{tab-3-resbisq4}
\end{table}

 Our work is "one
more" contribution to the study of this problem and brings some
answers, but also
 raises new questions.
Indeed, our results disagree with previous estimates, but we cannot
claim  for sure that our estimates are more reliable than those of
other authors. What is extremely clear is that the groups who
studied numerically universal combinations of amplitudes in the
$4-$state Potts model all noticed the extreme difficulty to take
into account properly the logarithmic terms. We believe that our
protocol is self-consistent in the sense that our criterion is to
obtain a relative stability of the correction to scaling
coefficients. The results that we report here, although a rough
estimate which calls for deeper analysis, reach a reasonable
confidence level. If this is indeed the case, one should identify
the reason of the discrepancy from the theoretical predictions of
Cardy and Delfino. In the conclusion, and in a footnote of one of
their papers, Delfino et al~\cite{DelfinoBarkemaCardy00} (p.533)
explain that their results are sensitive to the relative
normalization of the order and disorder operator form factors which
could be the origin of some troubles for the ratios
$\Gamma_+/\Gamma_L$ and $R_C$. This possible explanation seems
nevertheless to be ruled out (as mentioned by Enting and Guttmann
already) by the very good agreement between the theoretical
predictions and all numerical studies (both MC and SE) in the case
of the 3-state Potts model. Eventually let us mention that the
two-kink approximation used by Cardy and Delfino is exact for $q=2$
(Ising model) and quite good for $q=3$, but probably questionable
close to the marginal case $q\to 4$. As a conclusion, we are afraid
that this work opens more questions than it bring answers.

\section*{Acknowledgements}
\label{sec:7}
Discussions with A. Zamolodchikov, V. Plechko, W. Janke and M. Henkel, and
a correspondence with J. Cardy and J. Salas were very helpful.

LNS is grateful to the Statistical Physics group of the University
Henri Poincar\'e Nancy~1 and to the Theoretical group of the
University Milano--Bicocca for the kind hospitality. Financial
support from the twin research program between the Landau
Institute and the Ecole Normale Sup\'erieure de Paris and Russian
Foundation for Basic Research are also acknowledged.

%
%


\clearpage
\addcontentsline{toc}{section}{Index}
\flushbottom
\printindex

\end{document}